\documentclass[12pt]{article}
\begin{document}
\def\theequation{\arabic{section}.\arabic{equation}}
\newcommand{\be}{\begin{equation}}
\newcommand{\ee}{\end{equation}}
\begin{titlepage}
\title{The rotation of polarization by gravitational waves}
\author{Valerio Faraoni
\\{\small \it Physics Department, 
Bishop's University}\\ 
{\small \it 2600 College Street, Sherbrooke, Qu\'{e}bec, Canada 
J1M~0C8}\\
{\small email: vfaraoni@ubishops.ca}}
\date{}
\maketitle
\thispagestyle{empty}
\begin{abstract}
There are conflicting statements in the literature about the   
  gravitational Faraday rotation of the plane of 
   polarization of polarized electromagnetic radiation travelling 
   through a gravitational wave. This issue is reconsidered 
using  a simple formalism 
describing the rotation of the plane of polarization in  a 
gravitational field, in the geometric optics approximation.
It is shown that, to first order in the gravitational wave 
amplitude, the rotation  angle is a boundary effect which 
vanishes for localized (astrophysically generated) gravitational 
waves and  is  non-zero, but nevertheless negligible, for 
cosmological gravitational waves.
\end{abstract}
\vspace*{5truecm}
\begin{center}
{\bf PACS:} 04.30.-w, 04.30.Nk
\end{center}  
\begin{center}
{\bf Keywords:}  polarization, gravitational waves
\end{center}  

\end{titlepage}  \clearpage  \setcounter{page}{1}

\section{Introduction}
\setcounter{equation}{0}

There are many instances in astrophysics in which 
electromagnetic waves propagate through gravitational waves. The 
effects induced on a  light beam traversing 
gravitational waves of astrophysical or 
cosmological origin  include small deflections and frequency 
shifts, and they have been studied extensively in the literature 
(Wheeler 1960, 
Winterberg 1968, 
Zipoy 1966, 
Zipoy \& Bertotti 1968, 
Dautcourt 1969, 1974, 1975a,~b, 1977, 
Kaufmann 1970,
Bergmann 1975, 
Bertotti 1971,
Bertotti \& Catenacci 1975, 
Burke 1975,~1981,
Korotun 1970, 
Linder 1986,
McBreen \& Metcalfe 1988,
Allen 1989, 1990,
Braginsky {\em et al.} 1990,
Kovner 1990, Faraoni 1992, 1993, 1996, 1998, Fakir 
1993a,~b, 1994a,~b, 1996, 1997,
Pogrebenko {\em et al.} 1994, 1996,
Labeyrie 1993,
Frieman {\em et al.} 1994,
Durrer 1994,
Pyne {\em et al.} 1996,
Bar-Kana 1996,
Marleau \& Starkman 1996,
Bracco 1997,
Gwinn {\em et al.} 1997,
Kaiser \& Jaffe 1997,
Faraoni \& Gunzig 1998,
Bracco \& Teyssandier 1998,
Damour \& Esposito-Far\`{e}se 1998,
Kopeikin {\em et al.} 1999, 2006,
Larson \& Schild 2000,
Ragazzoni {\em et al.} 2003,
Lesovik {\em et al.} 2005,
Kopeikin \& Korobkov 2005). The problem of 
whether, and how 
much,  a 
gravitational wave rotates the 
plane of polarization of a polarized electromagnetic wave 
propagating  through it (gravitational Faraday rotation or 
Skrotskii effect) 
has been considered now and again in astrophysics 
(Faraoni 1993, Surpi \& Harari 1999, 
Cooperstock \& Faraoni 1993, 
Kopeikin \& Mashhoon 2002, Prasanna \& 
Mohanty 2002). It is interesting to 
determine whether this effect is  detectable with current 
or foreseeable technology in  realistic  astrophysical 
or cosmological situations, as a means of  detecting 
gravitational waves through their interaction with light from 
distant sources. The effect considered here is different 
from the polarization of the cosmic microwave background,  which 
is 
essentially Thomson scattering of photons of the cosmic 
microwave background by an anisotropic plasma, with the 
anisotropy caused by the shear associated with the 
gravitational wave. Instead, here we are interested in the 
{\em geometric} rotation of the plane of polarization caused 
directly 
by the presence of the 
gravitational wave. This effect was considered in  previous 
literature in the context of lensing by ``ordinary'' 
gravitational lenses, {\em i.e.}, localized mass distributions 
(Dyer \& Shaver 1992), and in this context the 
extension to 
non-conventional lenses, such as gravitational waves, was 
considered. It was found that, to first order in the 
gravitational wave amplitudes, the gravitational Faraday 
rotation is absent (Faraoni 1993, Cooperstock \& Faraoni 1993). 
This result has  implications for the observation of lensed 
polarized radio 
sources  (Kronberg {\em et al.} 1991). However, 
results that apparently 
contradict  this conclusion have since appeared in the 
literature (Surpi \& Harari 1999, Kopeikin \& 
Mashhoon 2002, Prasanna \& Mohanty 2002). The purpose of 
this note is  to study these discrepancies and to re-examine the 
validity of  the result of (Faraoni 1993). It is 
found 
that the 
different results of 
Surpi \& Harari 1999), Kopeikin \& 
Mashhoon 2002), and Prasanna \& Mohanty 2002) are due to the 
fact that different 
physical 
situations are studied, which is reflected in the different 
boundary conditions adopted. The first order systematic 
rotation of  the polarization plane described 
by Surpi \& Harari (1999) is a boundary effect 
which 
vanishes for localized gravitational waves.
Unfortunately, for cosmological gravitational waves which are 
not localized between the source and the observer, the effect is 
too  small to be observable. Moreover,   it  is  not a 
differential effect, which makes it undetectable if no 
independent information is available on the polarization of light
before propagation through gravitational waves. The 
effect reported in (Surpi \& 
Harari 1999, 
Kopeikin \& Mashhoon 2002 
and Prasanna \& Mohanty 2002) is too 
small  to be observed and it disappears in the limit $\lambda 
\ll  \lambda_{gw}$, where $\lambda$ and $\lambda_{gw}$ are the 
wavelengths of the  electromagnetic and gravitational waves, 
respectively.

\section{Polarized radiation crossing a gravitational wave}
\setcounter{equation}{0}

For simplicity, we consider a flat background 
described by the Minkowski metric $\eta_{\mu\nu}$, perturbed by 
localized gravitational waves. The resulting metric is 
$g_{\mu\nu}=\eta_{\mu\nu}+h_{\mu\nu}$ in an asymptotically 
Cartesian coordinate system $\left\{ x^{\alpha} \right\}$, with 
$\left| h_{\mu\nu} \right|\ll 
1$ and $h_{\mu\nu}\rightarrow 0 $ as $r\equiv \sqrt{ 
x^2+y^2+z^2 } \rightarrow +\infty$. In the eikonal  
approximation, the wavelength $\lambda $  of light propagating 
through the gravitational wave (of wavelength $\lambda_{gw}$) 
satisfies $\lambda \ll \lambda_{gw}$. The  
electromagnetic four-potential is 
\begin{equation} \label{1}
A_{\mu}=\hat{A}_{\mu} \left( x^{\alpha} \right) \, 
\mbox{e}^{i\omega S( x^{\alpha} )}  \;,
\end{equation}
where $\hat{A}_{\mu}$ is a slowly varying amplitude, 
$\omega$ is the angular frequency of the wave, and 
the rapidly varying eikonal $ S $ and its gradient 
$S_{\mu}\equiv \nabla_{\mu}S$ satisfy 
\begin{equation} \label{2}
S_{\mu} S^{\mu}=0 \;, \;\;\;\;\;\;\;\; A_{\mu} S^{\mu}=0 \;, 
\;\;\;\;\;\;\;\; S^{\nu} \nabla_{\nu} S^{\mu}=0 \;.
\end{equation}
It is assumed that 
\begin{equation} \label{3}
S^{\mu}={S^{(0)}}^{\mu} +\delta S^{\mu} = \left( 1,0,0,1 
\right)+\delta S^{\mu} \;,
\end{equation}
where $ {S^{(0)}}^{\mu} $ is the unperturbed tangent to the 
null geodesic, 
and $\delta S^{\mu}$ 
is a small perturbation induced by the gravitational wave. The 
perturbed geodesic equation yields ({\em e.g.}, Faraoni 1993)
\begin{equation} \label{4}
\delta S^{\mu}=-2\Big[ {h_0}^{\mu}+{h_3}^{\mu} \Big]_S^O 
+\frac{1}{2} \, \int_S^O dz \, \left( h_{00}+2h_{03}+h_{33} 
\right)^{, \mu} +\mbox{O}\left( 2 \right) \;,
\end{equation}
where ${}_{, \mu} $ denotes partial differentiation with 
respect to 
$x^{\mu}$. The quantities on the right hand side of 
eq.~(\ref{4}) are evaluated between the light source~S and the 
observer~O, and $\mbox{O}(2)$ denotes second order quantities 
in the gravitational wave amplitudes. The  first term on the 
right hand side is a boundary 
term and vanishes for astrophysically generated gravitational 
waves localized between the 
source and the observer. The amplitude of 
the electromagnetic potential is further decomposed as 
(Stephani 2004) $ \hat{A}^{\mu} \equiv a\, 
P^{\mu} $, 
where $a$ is a complex scalar and $P^{\mu}$ is a real vector 
satisfying (Stephani 2004)
\begin{eqnarray}
\frac{1}{a}\, \frac{da}{d\sigma} & = & - \, \theta\equiv 
-\frac{\nabla_{ \alpha} S^{\alpha}}{2} \;, \label{6} \\
&& \nonumber \\
\frac{dP^{\mu}}{d\sigma} &= &\frac{1}{2} \left( 
\frac{P^{\nu}\partial_{\nu}a}{a}+\nabla_{\nu}P^{\nu} 
\right)S^{\mu} \;. \label{7} 
\end{eqnarray}
Here $\sigma$ is an affine parameter along the null geodesic 
with tangent $S^{\mu}$ and  $ \theta $ is the 
expansion of a  congruence of null geodesics around a fiducial 
ray. The decomposition
\begin{eqnarray} 
a & = & a^{(0)}+\delta a=\frac{A}{\sigma}+\delta a \;, \label{8} 
\\
&& \nonumber \\
P^{\mu} &=& {P^{(0)}}^{\mu}+\delta P^{\mu}=\left( 0,1, 0,0
\right)+\delta P^{\mu} \label{9}
\end{eqnarray}
(where $A$ is a complex constant and ${P^{(0)}}^{\mu}$ 
corresponds to  
radiation polarized along the $x$-axis) yields
\begin{equation}
\frac{1}{ a^{(0)}}\, \frac{ d\left( \delta 
a\right)}{d\sigma}+\frac{1}{\sigma} \, \frac{ \delta 
a} {a^{(0)} } +\delta \theta =0 \;,
\end{equation}

\begin{equation}
\frac{ \left( \delta P^{\mu}\right)}{d\sigma}=\frac{1}{2} \left[ 
{P^{(0)}}^{\nu} \frac{ \partial_{\nu} \left( \delta a 
\right)}{a^{(0)} } +\nabla_{\nu}  P^{\nu}  
\right] {S^{(0)}}^{\mu} \;,
\end{equation}
where $\delta \theta = \theta-\theta^{(0)}$ is the perturbation 
of the expansion of a congruence of null geodesics. The 
four-divergence of $P^{\nu}$ is
\begin{equation}
\nabla_{\nu}  P^{\nu}=\frac{1}{\sqrt{-g}} \, 
\partial_{\mu}\left( \sqrt{-g} \, P^{\mu} 
\right)=\partial_{\mu}\left( \delta P^{\mu} \right)+\frac{1}{2} 
\, \partial_x h \;,
\end{equation}
where $h\equiv {h^{\mu}}_{\mu}$ and 
$\sqrt{-g}=1+h/2+\mbox{O}(2)$. As a result,
\begin{equation} \label{10}
\frac{ d \left( \delta P^{\mu} \right)}{d\sigma}=\frac{1}{2} 
\left[ \frac{ \partial_x \left( \delta a \right)}{a^{(0)} } 
+\partial_{\alpha}\left( \delta P^{\alpha} \right) +\frac{1}{2} 
\, \partial_x h \right] \left( \delta^{0\mu} +\delta^{3\mu} 
\right) +\mbox{O}(2) \;.
\end{equation}
Integration along the unperturbed photon path, instead of 
the perturbed one, only implies a second order error and yields
\begin{equation}\label{questa}
\delta P^{\mu}=\frac{1}{2} \left( \delta^{0\mu}+\delta^{3\mu} 
\right)\, \int_S^O dz \, \left[ \frac{\partial_x \left( \delta a 
\right)}{a^{(0)}} +\partial_{\alpha}\left( \delta P^{\alpha} 
\right) \right] +C^{\mu} + \mbox{O}(2) 
\end{equation}
in the transverse-traceless (TT) gauge in which $h=0$, and where 
$C^{\mu}$ are integration constants. 
Equation (\ref{questa}) yields $ \delta P^1  =\mbox{const.} $,  
$\delta P^2=\mbox{const.}$ 
When the boundary conditions $h_{\mu\nu}( S)=h_{\mu\nu}(O)=0$ 
describing localized gravitational waves are   
imposed,   $\delta P^1=\delta P^2=0$.  Because $P^{\mu}$ is a 
purely spatial vector, $\delta P^0=0$, which implies 
that the integral on the right hand side of eq.~(\ref{questa}) 
vanishes and then $\delta P^3 = $const. as 
well\footnote{Alternatively, one can rewrite eq.~(\ref{7}) as 
$dP^{\mu}/d\sigma = S^{\mu} \nabla^{\alpha}A_{\alpha} /(2a)$ and 
note that the right hand side vanishes in the Lorentz gauge 
$\nabla^{\mu}A_{\mu}=0$ and, therefore, in any gauge because 
this quantity is a scalar.}.

\section{Conclusions}
\setcounter{equation}{0}

The rotation angle is a 
boundary term effect which vanishes for 
intervening gravitational waves of astrophysical origin, which 
are localized between the source and the observer. Surpi \& 
Harari (1999), instead, consider a {\em 
cosmological}  gravitational wave, for which the boundary 
conditions consist of both $ h_{\mu\nu}(S) $ and $ 
h_{\mu\nu}(O) $  non-zero and $ h_{\mu\nu}(S) \neq  
h_{\mu\nu}(O) $. They find the 
rotation of the polarization vector of electromagnetic radiation 
(Surpi \& Harari 1999)
\begin{eqnarray}
&& \delta \theta = \frac{1}{2} \left( 1+\mu \right) \left[ 
h_{\times}\left( z_S, t_e 
\right)  -h_{\times} \left( 0, t_o \right) \right] 
\nonumber \\
&& +\frac{1}{4} 
\left( 1+\mu^2 \right)\Delta h_{+} \sin\left( 2\varphi \right) 
+\frac{\mu}{2} \Delta h_{\times} \cos\left( 2\varphi \right)
 \;, \label{12}
\end{eqnarray}
where $ \mu =\frac{ \vec{k}}{k}\cdot  \frac{ \vec{k}_{gw} 
}{k_{gw}}$, 
$\vec{k} $ is the electromagnetic wave vector, and $\vec{k}_{gw} 
$ is the gravitational wave vector. $t_e$ 
and $z_S$ are the emission time and the source position, while 
$t_O$ is the time at which the light is observed at the location 
of the observer $z=0$, and $h_{+}$ and $h_{\times}$ are the two 
independent polarizations of the gravitational wave in TT gauge.
Unperturbed light  propagates in 
the $z$-direction with wave  vector $\omega {S^{(0)}}^{\mu} $ 
which has the projection  $ \left( \cos\varphi, \sin \varphi \right) 
\sqrt{ 1-\mu} $ onto the $\left( x,y 
\right)$ plane. The effect described by 
eq.~(\ref{12}) is clearly an endpoint effect due entirely to 
the boundary conditions describing a cosmological gravitational 
wave and different from those corresponding to a localized wave 
employed in (Faraoni 1993).

While only the gauge-dependent  four-potential $A_{\mu}$ is 
discussed here, one can conclude that 
because  the gravitational wave induces no effect in $A^{\mu}$ 
to first order, to the same order there is no  effect also in 
the  (gauge-invariant) Maxwell field   $  F_{\mu\nu} = 
\nabla_{\mu}A_{\nu}-\nabla_{\nu}A_{\mu} $ derived 
from it.

A careful analysis of gravitomagnetic effects in the 
propagation of electromagnetic waves through time-dependent 
gravitational fields, including gravitational waves, is 
performed by Kopeikin \& Mashhoon (2002) by 
explicitly expanding the gravitational field in multipoles.   
The rotation of the plane of polarization by quadrupolar 
gravitational waves is entirely given by endpoint terms 
(see eq.~(138) of Kopeikin \& Mashhoon 2002). Similarly, 
Prasanna 
\& Mohanty (2002) consider the  
rotation of the plane 
of polarization of electromagnetic waves propagating parallel 
to  a gravitational wave, in the specific case of pulses emitted 
by a binary pulsar.  They find the dispersion relation
\begin{equation}
k^2=\omega^2 \left( 1\pm \frac{G\mu_r d^2 
\omega_{gw}^2}{3\omega^2 
r} \, \mbox{e}^{i\omega_{gw}\left( t-z\right) } \right) \;,
\end{equation}
where $\mu_r $  is the reduced mass of the binary and $d$ is a 
parameter describing the projected size of the orbit and related 
to the binary system semi-major axis 
(Prasanna \& Mohanty 2002). 
This leads again to a rotation angle of the polarization vector 
consisting of boundary terms (eq.~(19) of (Prasanna \& Mohanty 
2002). Moreover, the correction to the flat 
space dispersion relation 
$k^2=\omega^2$ is of the order $  \left( 
\frac{\omega_{gw}}{\omega} \right)^2 h_{\mu\nu} $, and it 
becomes 
exceedingly small in the limit $\lambda \ll \lambda_{gw}$. In 
the most optimistic case considered in 
(Prasanna \& Mohanty 2002), the rotation angle 
of the 
polarization vector is of the order of $10^{-8}$ radians or less, 
leaving little hope for detection.

Another situation in which the gravitational wave is not 
localized, but the boundary conditions determine $\delta 
P^{\mu}=0$ at the endpoints, occurs in laser interferometric 
detectors of gravitational waves. In  
(Cooperstock \& Faraoni 1993), laser 
interferometers are 
studied by explicitly computing the  perturbations of the 
Maxwell tensor by the gravitational wave. By contrast, in most 
of the  literature on the subject, the phase shift between two 
different arms of the interferometer is computed by 
considering different travel times or different lengths travelled 
without explicitly  considering the changes induced in the 
electromagnetic field, and in the approximation $\lambda 
\ll \lambda_{gw}$.  
In an interferometer's arm, an electromagnetic wave is not 
localized between a ``source'' and an ``observer'', but it spans 
the 
entire  length of the arm. In  
(Cooperstock \& Faraoni 1993),  
no rotation of the polarization  vector of electromagnetic 
radiation was found, to first order in  the gravitational wave 
amplitude. This fact is again explained  by the boundary 
conditions imposed at the endpoints: at these 
locations, reflection off perfect mirrors is assumed and the 
electromagnetic field describing a standing wave between the 
two mirrors, which are nodes, also vanishes. As a consequence, 
the endpoint 
effect vanishes too.

As a  conclusion, the discrepancy between 
(Faraoni 1993) 
and (Surpi \& Harari 1999, Kopeikin \& Mashhoon  2002, and 
Prasanna \& Mohanty 2002) is due 
to the 
different boundary conditions. The rotation angle always 
vanishes, to first order, for localized gravitational waves and 
is always reducible to an endpoint effect in 
(Surpi \& Harari 1999, Kopeikin 
\& Mashhoon 2002, Prasanna \& 
Mohanty 2002). 
Unfortunately,  the detection of gravitational waves through 
the rotation of the plane of polarization of light from 
distant  sources is not feasible with technology currently 
available  or foreseeable in the near future. On the other hand, 
we should  not worry about the plane of polarization of 
polarized radiation  emitted from radio galaxies 
(Kronberg {\em et al.} 1991) being altered by 
gravitational waves along 
the line of sight,  in the same way that  gravitational 
Faraday rotation is  negligible in weak lensing by ordinary 
gravitational lenses (Dyer \& Shaver 1992, Faraoni 1993).

\section*{Acknowledgments}
 
This work was supported by the Natural Sciences and Engineering  
Research Council of Canada  (NSERC).

\end{document}